\documentclass[11pt,fleqn]{article} 

\usepackage[utf8]{inputenc} 

\usepackage{fullpage} 
\usepackage{graphicx,color}
\usepackage{bm}
\usepackage{bbm}
\usepackage{amsmath,amssymb,amsfonts,amsgen,amsbsy,units}
\usepackage{booktabs} 
\usepackage{array} 
\usepackage{paralist} 
\usepackage{verbatim} 
\usepackage{subfig} 
\usepackage[numbers, sort&compress]{natbib}

\usepackage{hyperref}
\hypersetup{colorlinks=true, linkcolor=blue, citecolor=bluelinkcolor=blue, citecolor=blue}

\providecommand\bnabla{\bm{\nabla}}
\providecommand\der{\mathrm{d}}
\renewcommand{\vec}[1]{\ensuremath{\bm{#1}}} 
\providecommand\bnabla{\gvec{\nabla}}
\renewcommand{\cdot}{\;\vcenter{\hbox{\tiny$\bullet$}}\;} 

\newcommand{\aderv}[2]{\frac{\partial {#1}}{\partial {#2}}}
\newcommand{\adervo}[2]{\frac{\der {#1}}{\der {#2}}}

\newcommand{\adervso}[2]{\frac{\der^2 {#1}}{\der {#2}^2}}

\newcommand{\equa}[1]{Eq.~(\ref{#1})}

\newcommand{\equas}[1]{Eqs.~(\ref{#1})}
\newcommand{\equass}[2]{Eqs.~(\ref{#1})--(\ref{#2})}
\newcommand{\equasa}[2]{Eqs.~(\ref{#1}){ }and{ }(\ref{#2})}

\newcommand{\eqn}[2]{\begin{gather}
\displaybreak[2]
#1
\label{#2}
\end{gather}
}

\newcommand{\gat}[2]{\begin{subequations}\label{#2}\begin{gather}
#1
\end{gather}\end{subequations}
}

\title{\bf Buoyant flow and instability in a vertical cylindrical porous slab with permeable boundaries}
\author{{\bf A. Barletta}$^1$\footnote{Email address for correspondence: {\tt antonio.barletta@unibo.it}}\ ,\ {\bf M. Celli}$^1$,\ {\bf D.A.S. Rees}$^2$\\[3mm]
$^1${\small Department of Industrial Engineering, Alma Mater Studiorum Universit\`a di Bologna,}\\
{\small Viale Risorgimento 2, 40136 Bologna, Italy}\\[3mm]
$^2${\small Department of Mechanical Engineering, University of Bath,}\\
{\small Claverton Down, Bath BA2 7AY, United Kingdom}\\
}

\date{} 

\begin{document}
\maketitle

\begin{abstract}\noindent
The basic stationary buoyant flow in a vertical annular porous passage induced by a boundary temperature difference is investigated. The vertical cylindrical boundaries are considered both isothermal and permeable to external fluid reservoirs. There exists a stationary parallel velocity field with a zero flow rate and pure conduction heat transfer. Its linear stability is analysed with normal mode perturbations of the pressure and temperature fields. The transition to convective instability is caused by the basic horizontal temperature gradient. Hence, its nature differs from that of the usual Rayleigh--B\'enard instability. The linear dynamics of the perturbed flow is formulated as an eigenvalue problem, solved numerically. Its solution provides the neutral stability curve at each fixed aspect ratio between the external radius and the internal radius. The critical Rayleigh number triggering the instability is evaluated for different aspect ratios. It is shown that the system becomes more an more unstable as the aspect ratio increases, with the critical Rayleigh number dropping to zero when the aspect ratio tends to infinity.\\[5mm]
\end{abstract}

\newpage

\section{Introduction}
There is a wide research area in heat transfer regarding the onset conditions for convection cells either in fluids or in fluid saturated porous media. Several studies have been published within this area which are focussed, mainly, on the emergence of the Rayleigh--B\'enard instability and its many variants \citep{drazin2004hydrodynamic}. The distinctive feature of the Rayleigh--B\'enard instability relies on the vertical, downward oriented, temperature gradient eventually causing the development of convective cells in an otherwise thermally stratified fluid. Such an instability can be developed either when the fluid is at rest or in the presence of a basic fluid flow. The Rayleigh--B\'enard instability, when it takes place in a fluid saturated porous medium, is termed either Horton--Rogers--Lapwood instability or Darcy--B\'enard instability \citep{rees2000stability, straughan2008stability, nield2017convection, barletta2019routes}. 

The thermal instability is not necessarily of the Rayleigh--B\'enard type, as it may arise with a basic temperature gradient inclined to the vertical. This is the case of the Hadley flow, a phenomenon of paramount importance in the field of atmospheric physics \citep{hart1972stability, weber1973thermal, sweet1977free}. With reference to porous media, the Hadley flow may also give rise to an unstable behaviour as pointed by several authors \citep{weber1974convection, daniels1986thermally, daniels1989thermally, nield1991convection, nield1993convection,  barletta2010instability, barletta2012hadley, barletta2012linear}. Another instance where the thermal instability departs from the Rayleigh--B\'enard type happens with a vertical plane layer of fluid, or of saturated porous material, bounded by isothermal planes with different temperatures \citep{vest1969stability, gill1969proof}. With such conditions, the basic parallel flow is coupled with a purely horizontal temperature gradient.
A remarkable difference between the case of a fluid in a vertical slot and that of a vertical porous layer is that the former case allows the emergence of a thermal instability \citep{vest1969stability}, while no instability is possible for the latter case \citep{gill1969proof}. A recent study \citep{barletta2015proof} made it clear that the lack of instability in a vertical porous slab proved by \citet{gill1969proof} relies dramatically on the assumption that the bounding planes of the slab are impermeable. On relaxing this assumption, a thermal instability emerges when the Rayleigh number is larger than $197.081$ \citep{barletta2015proof}. The boundary conditions assumed by \citet{barletta2015proof} are of permeable open boundaries.  This analysis was further extended by considering conditions of local thermal non--equilibrium in the saturated porous medium \citep{celli2017local}.

The aim of this paper is a development of the results presented by \citet{barletta2015proof} by envisaging the case where, instead of a vertical plane slab, we consider a vertical cylindrical slab. The focus is on the role of the curvature, parametrised through the aspect ratio between the external and internal radii, on the instability. A linear stability analysis will be carried out aimed to obtain the neutral stability conditions and the values of the Rayleigh number, wavenumber and angular frequency for the critical modes which activate the instability. The limiting case of a plane slab, described by \citet{barletta2015proof}, is obtained when the aspect ratio tends to unity. The substantial role of the curvature induces a destabilisation of the basic parallel flow as the aspect ratio increases. Ultimately, the basic parallel flow is always unstable when the aspect ratio tends to infinity. This limiting case is that of a semi--infinite porous medium bounded internally by a cylindrical surface.

\begin{figure}[t]
\centering
\includegraphics[width=0.4\textwidth]{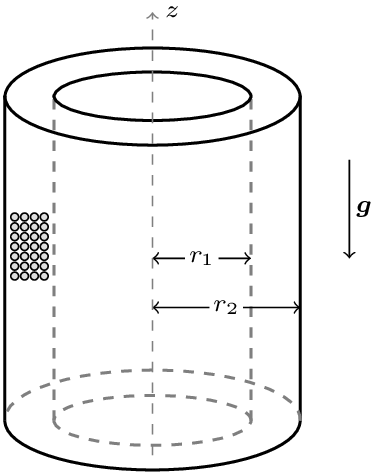}
\caption{{A sketch of the porous cylindrical slab}}
\label{fig1}
\end{figure}

\section{Mathematical model}
We study the buoyant flow in a vertical porous layer saturated by a fluid and bounded by two coaxial cylindrical surfaces with radii $r_1$ and $r_2$. Cylindrical coordinates $(r, \phi, z)$ are adopted under the assumption that the annular layer has an infinite length in the vertical $z$ direction. Figure~\ref{fig1} shows a sketch of the vertical porous layer. The gravitational acceleration $\vec{g}$, with modulus $g$, 
{acts in the downward vertical direction.}

The boundaries $r=r_1$ and $r=r_2$ are supposed to be isothermal with given temperatures $T_1$ and $T_2$, respectively. Such boundaries are permeable to an external fluid environment at rest. The permeable boundary conditions are modelled with a uniform pressure which coincides with the hydrostatic pressure \citep{barletta2015proof}.

By adopting Darcy's law within the framework of the Oberbeck--Boussinesq approximation, we can write the governing equations expressing the local mass, momentum and energy balance in the fluid--saturated porous medium in a dimensionless formulation, namely
\gat{
\bnabla \cdot \vec{u} = 0 , \label{1a}\\
\vec{u} = -\, \bnabla p + R\, T \, \vec{e}_z, \label{1b}\\
\aderv{T}{t} + \vec{u} \cdot \bnabla T = \nabla^2 T , \label{1c}
}{1}
where $\vec{u}$ is the seepage velocity having components $(u,v,w)$ in the radial, angular and axial directions, respectively. Moreover, $p$ is the local difference between the pressure and the hydrostatic pressure, $t$ is the time, $T$ is the temperature and $\vec{e}_z$ is the unit vector along the $z$ axis. 
The parameter $R$ is the porous medium definition of the Rayleigh number \citep{nield2017convection}, given by
\eqn{
R = \frac{g \beta (T_2 - T_1) K r_1}{\nu \alpha} ,
}{2}
while the dimensionless coordinates and fields have been defined through the scaling
\eqn{
\frac{1}{r_1}\ (r, z) \to (r, z),\qquad \frac{\alpha}{\sigma r_1^2}\ t \to t, \qquad \frac{K}{\mu \alpha}\ p \to p,\nonumber\\
\frac{r_1}{\alpha}\ \vec{u} = \frac{r_1}{\alpha}\ (u, v, w) \to (u, v, w) = \vec{u},\qquad  \frac{T - T_0}{T_2 - T_1} \to T.
}{3}
Here, $\alpha$ is the average thermal diffusivity and $K$ is the permeability of the porous medium, $\mu$ and $\nu$ are the dynamic and kinematic viscosity coefficients of the fluid, $\beta$ is the thermal expansion coefficient of the fluid, while $\sigma$ is the ratio between the average volumetric heat capacity of the porous medium and that of the fluid. Finally, $T_0$ is the reference fluid temperature employed when formulating the Oberbeck--Boussinesq approximation and the definition of the buoyancy force.

The dimensionless boundary conditions completing the mathematical model described by \equas{1} can be written as
\eqn{
r=1 : \qquad p = 0, \quad T = \zeta - 1, \nonumber\\
r=s : \qquad p = 0, \quad T = \zeta, 
}{4}
where
\eqn{
s=\frac{r_2}{r_1}, \qquad \zeta = \frac{T_2 - T_0}{T_2 - T_1} .
}{5}
Fixing a numerical value of $\zeta$ means deciding which reference temperature $T_0$ is more appropriate for optimising the Oberbeck--Boussinesq approximation and how $T_0$ depends on $T_1$ and $T_2$. We will return on this point later on.
 

\section{The basic buoyant flow}
From \equasa{1}{4} steady state parallel flow exists which is directed along the $z$ axis and given by
\eqn{
u_b = 0, \qquad v_b =0, \qquad w_b = R\, T_b, \qquad p_b = 0, \nonumber\\
T_b = \frac{2 \ln(r) + 1}{2 \ln(s)} - \frac{1}{s^2 - 1} - 1,
}{7}
where the subscript ``$b$'' in \equa{7} is for basic solution. This solution implies a specific choice for the parameter $\zeta$, namely
\eqn{
\zeta = \frac{1}{2 \ln (s)} - \frac{1}{s^2 - 1},
}{8}
or, equivalently, the reference temperature $T_0$ is chosen as the mean temperature across the annular cross--section. In fact, by using \equa{7}, one can easily check that the condition 
\eqn{
0 = \frac{1}{\pi (s^2 - 1)} \int_1^s T_b \; 2 \pi r\, \der r = \frac{2}{s^2 - 1} \int_1^s T_b \, r\, \der r
}{9}
is satisfied. Equations~(\ref{3}) and (\ref{9}) prove that $T_0$ is the mean temperature across the annular cross--section. As pointed out by \citet{barletta1999choice}, this is the optimal choice for the reference temperature to be employed within the Oberbeck--Boussinesq approximation. As $w_b = R\,T_b$, we note that the choice expressed by \equa{8} yields not only a zero average value of $T_b$, but also of $w_b$. This means that the basic vertical flow rate is zero.
%

\section{Linear stability}
The linear stability of the basic solution, \equa{7}, is tested by perturbing the basic flow with small amplitude disturbances, namely
\eqn{
u = u_b + \varepsilon \, \hat u , \qquad v = v_b + \varepsilon \, \hat v , \qquad w = w_b + \varepsilon \, \hat w, \nonumber\\
p = p_b + \varepsilon \, \hat p  , \qquad T = T_b + \varepsilon \, \hat T ,
}{10}
where $\varepsilon$ is the perturbation parameter and $|\varepsilon| \ll 1$. The substitution of \equa{10} into \equasa{1}{4}, by neglecting terms $O(\varepsilon^2)$ and by taking into account \equa{7}, yields
\gat{
 \frac{1}{r}\, \aderv{(r\, \hat u)}{r} + \frac{1}{r}\, \aderv{\hat v}{\phi} + \aderv{\hat w}{z}= 0 , \label{11a}\\
\hat{u} = - \aderv{\hat p}{r}, \label{11b}\\
\hat{v} = -\, \frac{1}{r}\; \aderv{\hat p}{\phi} , \label{11c}\\
\hat{w} = - \aderv{\hat p}{z} + R\, \hat{T} , \label{11d}\\
\aderv{\hat T}{t} + R\, T_b \; \aderv{\hat T}{z} + \hat{u}\; \adervo{T_b}{r} = \nabla^2 \hat{T} . \label{11e}
}{11}

\subsection{Pressure--temperature formulation}
If one prescribes a zero divergence of the perturbation velocity $(\hat{u}, \hat{v}, \hat{w})$, namely if one substitutes \equass{11b}{11d} into \equa{11a}, then a pressure--temperature formulation is attained given by
\gat{
\nabla^2 \hat{p} = R\, \aderv{\hat{T}}{z} , \label{12a}\\
\aderv{\hat T}{t} + R\, T_b \; \aderv{\hat T}{z} - \aderv{\hat p}{r}\; \adervo{T_b}{r} = \nabla^2 \hat{T} . \label{12b}
}{12}
On account of \equasa{7}{10}, \equa{4} yields the boundary conditions for the perturbations,
\eqn{
r=1,s : \qquad \hat{p} = 0, \quad \hat{T} = 0. 
}{13}

\subsection{Normal modes}
General three-dimensional normal modes are given by
\eqn{
\hat{p}(r,\phi,z,t) = f(r)\, \cos(m\,\phi)\, e^{i(k z - \omega t)}\, e^{\eta t}, \nonumber\\
\hat{T}(r,\phi,z,t) = h(r)\, \cos(m\,\phi)\, e^{i(k z - \omega t)}\, e^{\eta t}, \qquad m=0,1,2,\ \ldots\ ,
}{14}
where $k$ is the wavenumber, $\omega$ is the angular frequency and $\eta$ is the growth rate. The substitution of \equa{14} into \equasa{12}{13} yields the eigenvalue problem
\gat{
f'' + \frac{1}{r}\; f' - \left( \frac{m^2}{r^2} + k^2 \right) f - i\,k\, R\, h = 0 , \label{15a}\\
h'' + \frac{1}{r}\; h' - \left( \frac{m^2}{r^2} + k^2 + i\, k\, R\, T_b + \lambda \right) h + f'\, T_b' = 0, \label{15b}\\
r=1,s : \qquad f = 0, \quad h = 0, \label{15c}
}{15}
where $\lambda = \eta - i\, \omega$ {is the complex growth rate}
and primes denote derivatives with respect to $r$. 
For prescribed input parameters $(s,m,k,R)$, the solution of \equas{15} yields the eigenfunction pair $(f,h)$ together with the eigenvalue $\lambda$. The sign of the real part of $\lambda$, that is the growth rate $\eta$, allows one to detect the stable/unstable nature of the normal modes. The continuous transition of $\eta$ from a negative to a positive value entails the onset of the instability, with $\eta = 0$ yielding the neutral stability condition.

\subsection{Solution method}\label{solmed}
There are several approaches to the numerical solution of stability eigenvalue problems. Thorough presentations of the main features of numerical methods for differential eigenvalue problems are available, for instance, in \citet{dongarra1996chebyshev, straughan1996two}, chapter 9 of \citet{straughan2008stability}, as well as in chapter 10 of the recent book by \citet{barletta2019routes} though restricted to the shooting method. 

In our study, the shooting method is employed. The main steps for its implementation are: 
\begin{itemize}
\item The formulation of an initial value problem issued from \equas{15} by suitably extending the conditions set up at $r=1$. This means imposing
\eqn{
r=1 : \qquad f = 0, \quad f' = \xi_1 + i\, \xi_2, \quad h = 0, \quad h' = 1, 
}{16}
where $(\xi_1, \xi_2)$ are unknown real parameters defined in order to prescribe the initial condition for $f'$. The constraint $h=1$ is the exploitation of the scale fixing for the eigenfunctions allowed as a consequence of the homogeneous nature of \equas{15}.
\item The end conditions,
\eqn{
r=s : \qquad f = 0, \quad h = 0, 
}{17}
are used to determine the four real unknowns $(\xi_1, \xi_2, \eta, \omega)$. The consistency is ensured by the complex--valued nature of the eigenfunction pair, $(f,h)$. 
\end{itemize}
The solution procedure has to be applied for every assignments of the input data $(s,m,k,R)$. 
 
The shooting method is implemented via the {\sl Mathematica} engine (\copyright{} Wolfram Research, Inc.). To be more specific, functions {\tt NDSolve} and {\tt FindRoot} are employed for the solution of the initial value problem by prescribing the end conditions at $r=s$ given by \equa{15c}. The numerical computation of the eigenvalue $\lambda=\eta-i \, \omega$ allows one to determine the neutral stability condition where $\eta=0$. This condition occurs with values of $R$ and $\omega$ that will be computed for prescribed $s$, $m$ and $k$. 

\begin{table}[t]
\begin{center}
\begin{tabular}{l|c|c}
Method & $R$ & $\omega$\\
\hline\hline
 & & \\[-3mm]
$\delta r=10^{-1}$ & 390.3442641 & $-26.73827064$ \\
$\delta r=10^{-2}$ & 393.8312951 & $-27.31044536$ \\
$\delta r=10^{-3}$ & 393.8316177 & $-27.31047427$ \\
$\delta r=10^{-4}$ & 393.8316177 & $-27.31047427$ \\
adaptive           & 393.8316113 & $-27.31047363$ \\\end{tabular}
\end{center}
\caption{\label{tab1}Comparison between fixed step--size algorithm, with a given $\delta r$, and the adaptive step--size algorithm for the numerical computation of the neutral stability $(\eta=0)$ values of $R$ and $\omega$ with $s=1.5$, $m=0$, $k=2$}
\end{table}%

Table~\ref{tab1} reports the numerical values of $R$ and $\omega$ at the neutral stability threshold $(\eta=0)$ for the sample case $s=1.5$, $m=0$, $k=2$. The numerical solution is carried out either by using an explicit fourth--order Runge--Kutta solver with a fixed step--size, $\delta r$, or with an adaptive step--size algorithm. For the fixed step--size algorithm, gradually decreasing values of $\delta r$ are assigned. The lowest threshold value of $R$ for the transition from a negative $\eta$ to a positive $\eta$ is detected. The agreement between the two solution strategies is excellent with, at least, seven significant figures matching perfectly. The advantage of the adaptive step--size method is in its lower computational time with respect to the fixed step--size method for either $\delta r=10^{-4}$ or $\delta r=10^{-3}$. Thus, all the numerical data discussed in the forthcoming sections will be obtained by employing the adaptive step--size method.

\subsection{Symmetry}
An important symmetry of the stability eigenvalue problem correspond to changing the sign of $R$ which, physically, means that the heated boundary switches from $r = s$ to $r = 1$ or vice versa. One may recognise that {this modification to the original} eigenvalue problem may be solved by taking the complex conjugate of \equas{15}, namely
\gat{
\bar{f}'' + \frac{1}{r}\; \bar{f}' - \left( \frac{m^2}{r^2} + k^2 \right) \bar{f} + i\,k\, R\, \bar{h} = 0 , \label{18a}\\
\bar{h}'' + \frac{1}{r}\; \bar{h}' - \left( \frac{m^2}{r^2} + k^2 - i\, k\, R\, T_b + \bar{\lambda} \right) \bar{h} + \bar{f}'\, T_b' = 0, \label{18b}\\
r=1,s : \qquad \bar{f} = 0, \quad \bar{h} = 0, \label{18c}
}{18}
where the overline denotes complex conjugation.
By comparing \equas{15} with \equas{18}, we conclude that the two eigenvalue problems agree by transforming \equas{18} with
\eqn{
(\bar{f},\bar{h}) \to (f,h), \qquad R \to -R \qquad \bar{\lambda} \to \lambda .
}{19}
One can infer from \equa{19} that, for given values of $(s,m,k)$, the growth rate $\eta$ corresponding to a prescribed $R$ coincides with that corresponding to $-R$, while the angular frequency $\omega$ changes its sign. In other words, the neutral stability curves drawn in the $(k, |R|)$ plane are the same for either positive or negative $R$.

As a consequence of this symmetry, one can conclude that it is not restrictive to base the forthcoming analysis on the assumption of non--negative values of $R$.

\begin{figure}
\centering
\includegraphics[width=\textwidth]{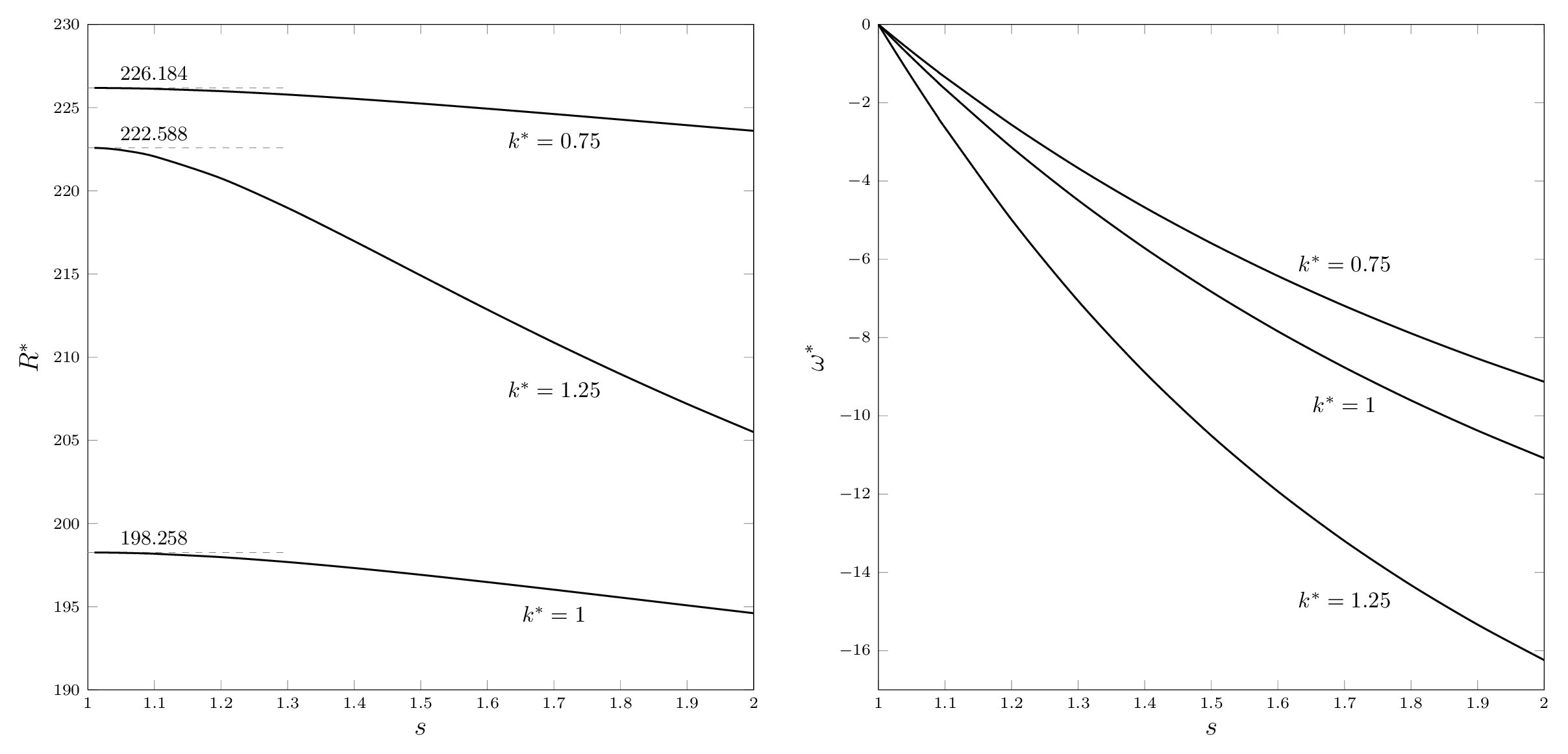}
\caption{Neutral stability data for axisymmetric modes $(m=0)$: plots of $R^*$ and $\omega^*$ versus $s$ for different values of $k^*$. The asymptotic values of $R^*$ for the plane slab case \citep{barletta2015proof} are reported for comparison with the limit $s \to 1$}
\label{fig2}
\end{figure}

\section{Discussion of the results}
The analysis of the numerical data for neutral stability is presented by first inspecting the limiting case where the effects of curvature become negligible. This happens when $s \to 1$. Then, the general case is described by fixing different values of the aspect ratio $s$.

\subsection{Asymptotic case $s \to 1$}
In order to retrieve the neutral stability condition for the case where the effects of curvature can be neglected, we must consider a cylindrical slab where the ratio $s$ is very close to $1$. We expect to match the analysis presented by \citet{barletta2015proof} in this asymptotic case. This limit can be taken consistently by first recognising that the dimensionless variables defined here are different from those defined for the plane slab. The discrepancy is due to the different reference lengths adopted, {\em i.e.} the inner radius $r_1$ here and the layer thickness for the plane slab. This affects the definition of the dimensionless coordinates, time and velocity, as well as the definition of the Rayleigh number and, indirectly, the definitions of $k$ and $\lambda$. In fact, the latter quantities are, dimensionally, an inverse length and an inverse time, respectively. Therefore, we introduce alternative definitions of $R$, $k$ and $\lambda$ as follows:
\eqn{
R^* = (s-1) R, \qquad k^* = (s-1) k, \qquad \lambda^* = \eta^* - i \omega^* = (s-1)^2 \lambda.
}{20}
The evaluation of the neutral stability data shows that $R$, $k$ and $\omega$ tend to infinity when $s \to 1$ in such a way that $R^*$, $k^*$ and $\omega^*$ attain a finite limit.

\begin{figure}
\centering
\includegraphics[width=0.5\textwidth]{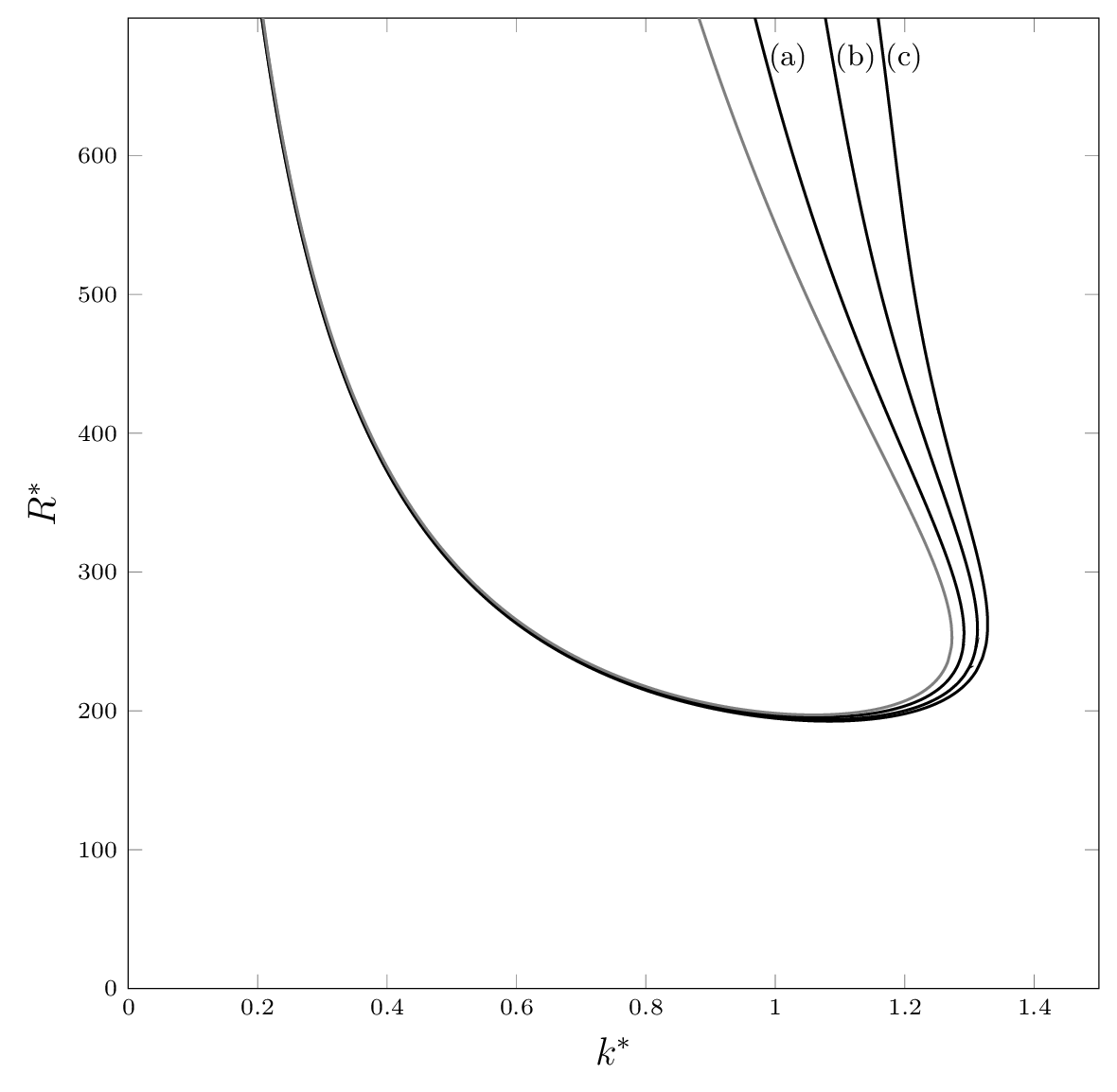}
\caption{Neutral stability curves for axisymmetric modes $(m=0)$ in the $(k^*, R^*)$ parametric plane. The black curves are for: (a) $s=1.5$; (b) $s=1.8$; (c) $s=2$. The grey line is relative to the plane slab case \citep{barletta2015proof}, namely the asymptotic case $s \to 1$}
\label{fig3}
\end{figure}

Figure~\ref{fig2} illustrates the change in the neutral stability values of $R^*$ and $\omega^*$ as the aspect ratio $s$ tends to unity. Three different wavenumbers $k^*$ are monitored with the neutral stability condition, {\em i.e.} $\eta=0$, traced for axisymmetric modes $(m=0)$. The expected plane slab limit is attained when $s \to 1$. This is revealed in the left hand frame of Fig.~\ref{fig2} by comparison with the neutral stability data found for the plane slab problem \citep{barletta2015proof}. A particularly evident feature is that $\omega^*$ tends to $0$ for every $k^*$ when $s \to 1$. This means that the non--travelling nature of the neutrally stable modes is retrieved in the plane slab limit, even if the transition to convective instability happens with travelling modes $(\omega^* \ne 0)$ for every $s > 1$. This result is a consequence of the curvature or, equivalently, of the different areas of the heated and cooled boundaries.

Some neutral stability curves are drawn in the parametric plane $(k^*, R^*)$ for $s=1.5, 1.8$ and $2$. They are reported in Fig.~\ref{fig3} together with the plane slab asymptotic case $(s\to 1)$, for comparison. As already pointed while commenting on Fig.~\ref{fig2}, the limit $s\to 1$ is approached continuously. Another important feature gathered from Fig.~\ref{fig3} is that the point with minimum $R^*$ along the neutral stability curve, {\em i.e.} the critical point $(k_c^*, R_c^*)$, yields a critical value of the Rayleigh number, $R_c^*$, which decreases slowly with $s$. These findings suggest that the curvature of the annular slab induces a destabilisation of the basic state.

\begin{figure}[t!!]
\centering
\includegraphics[width=\textwidth]{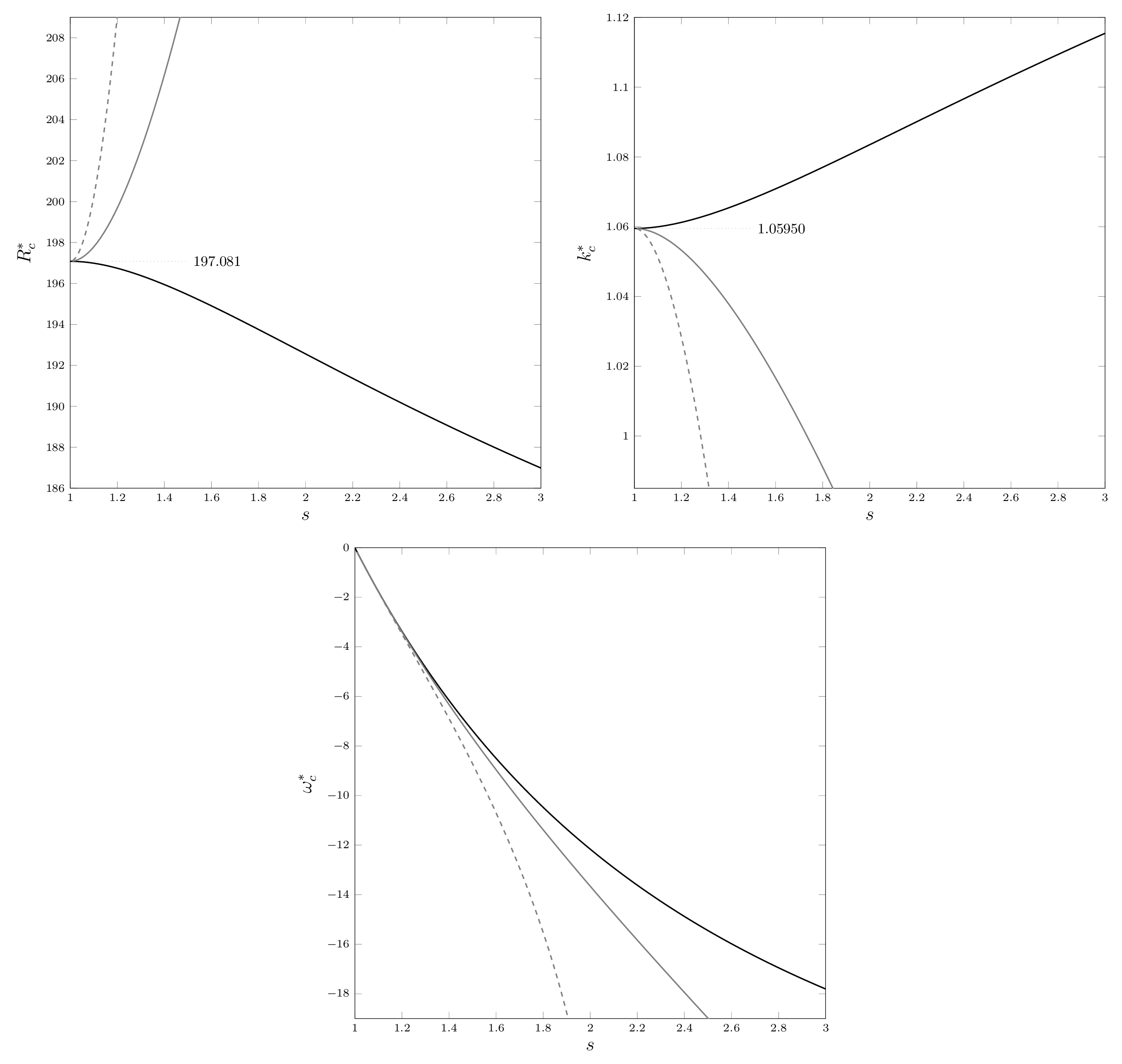}
\caption{Critical values of $R^*$, $k^*$ and $\omega^*$. Black solid lines are for axisymmetric modes, $m=0$. Gray solid lines are for $m=1$ modes, while gray dashed lines are for $m=2$ modes. The critical values relative to the plane slab case \citep{barletta2015proof}, {\em i.e.} the limit $s \to 1$ are reported for comparison}
\label{fig4}
\end{figure}

The behaviour of axisymmetric $(m=0)$ modes and non--axisymmetric modes with either $m=1$ or 2 is illustrated in Fig.~\ref{fig4} where the critical values of $R_c^*$, $k_c^*$ and $\omega_c^*$ are plotted versus $s$ in the range $1 < s \leqslant 3$. The most evident feature is that the axisymmetric modes yield the lowest values of $R_c^*$ and, hence, they are the most unstable. 

The destabilising effect of an increasing aspect ratio $s$ is confirmed by the trend of the critical value of $R^*$ for $m=0$, reported in Fig.~\ref{fig4}. An interesting aspect is that the behaviour for the limiting case $s \to 1$, {\em i.e.} the special case where the cylindrical slab tends to the plane slab, the critical values of $R_c^*$, $k_c^*$ and $\omega_c^*$ become independent of the angular number $m$. 

\begin{figure}[t!!]
\centering
\includegraphics[width=\textwidth]{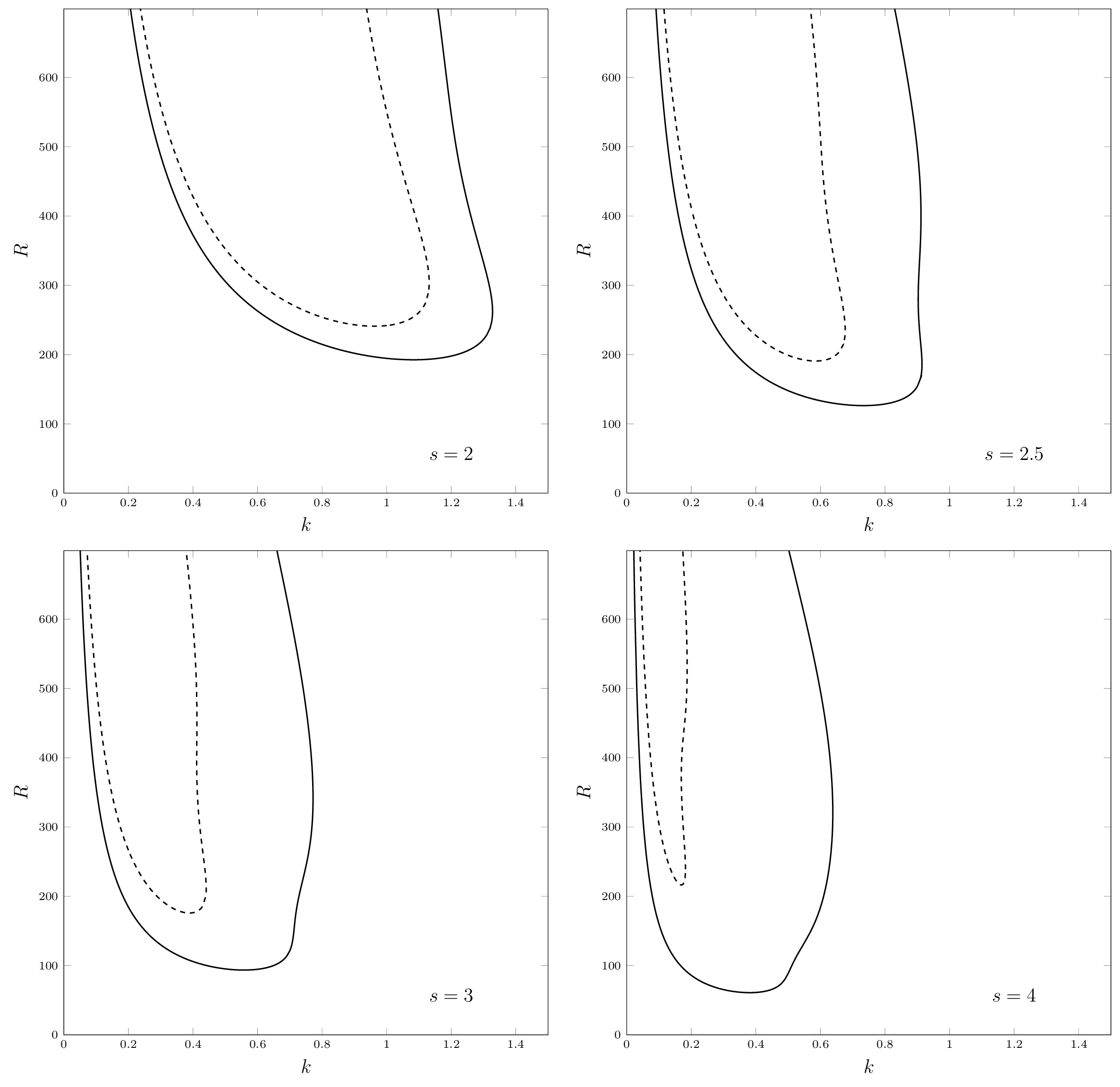}
\caption{Neutral stability curves in the $(k, R)$ parametric plane for axisymmetric modes $(m=0)$ and for $m=1$ non-axisymmetric modes, with $s=2,\ 2.5,\ 3$ and $4$. The solid curves are for $m=0$, while the dashed curves are for $m=1$}
\label{fig5}
\end{figure}

\subsection{Normal modes with $k=0$}
We can easily inspect the solution of \equas{15} when $k=0$. First, we note that \equa{15a} is simplified to
\eqn{
f'' + \frac{1}{r}\; f' - \frac{m^2}{r^2} \; f = 0 .
}{21}
The solution of \equa{21} satisfying the boundary condition $f(1)=0$ is given by
\eqn{
f(r) = A \, \sinh[m \ln(r)] .
}{22}
Then, the end boundary condition $f(s)=0$ can be satisfied only with $A=0$ or $m=0$. In either cases, the consequence is $f(r)=0$ for every $r$. We can rewrite \equa{15b} by  employing a change of the independent variable, {\em i.e.} by defining $x = \ln(r)$,
\eqn{
\adervso{h}{x} - \left( m^2 + \lambda\, e^{2 x} \right) h = 0 .
}{23}
We multiply \equa{23} by the complex conjugate of $h$ and we integrate by parts over the interval $0 \leqslant x \leqslant \ln(s)$. If we employ the boundary conditions~(\ref{15c}), we obtain
\eqn{
\int\limits_0^{\ln(s)}\; \left| \adervo{h}{x} \right|^2 \der x + m^2 \int\limits_0^{\ln(s)} \left| h \right|^2 \der x + \lambda \int\limits_0^{\ln(s)} e^{2 x} \left| h \right|^2 \der x = 0 .
}{24}
Thus, by recalling that $\lambda = \eta - i\, \omega$, we can infer that \equa{24} cannot be satisfied with $\eta > 0$ unless $h$ is identically zero. The conclusion is that no linear instability is possible with axially invariant modes $(k=0)$.

\begin{figure}[t!!]
\centering
\includegraphics[width=0.5\textwidth]{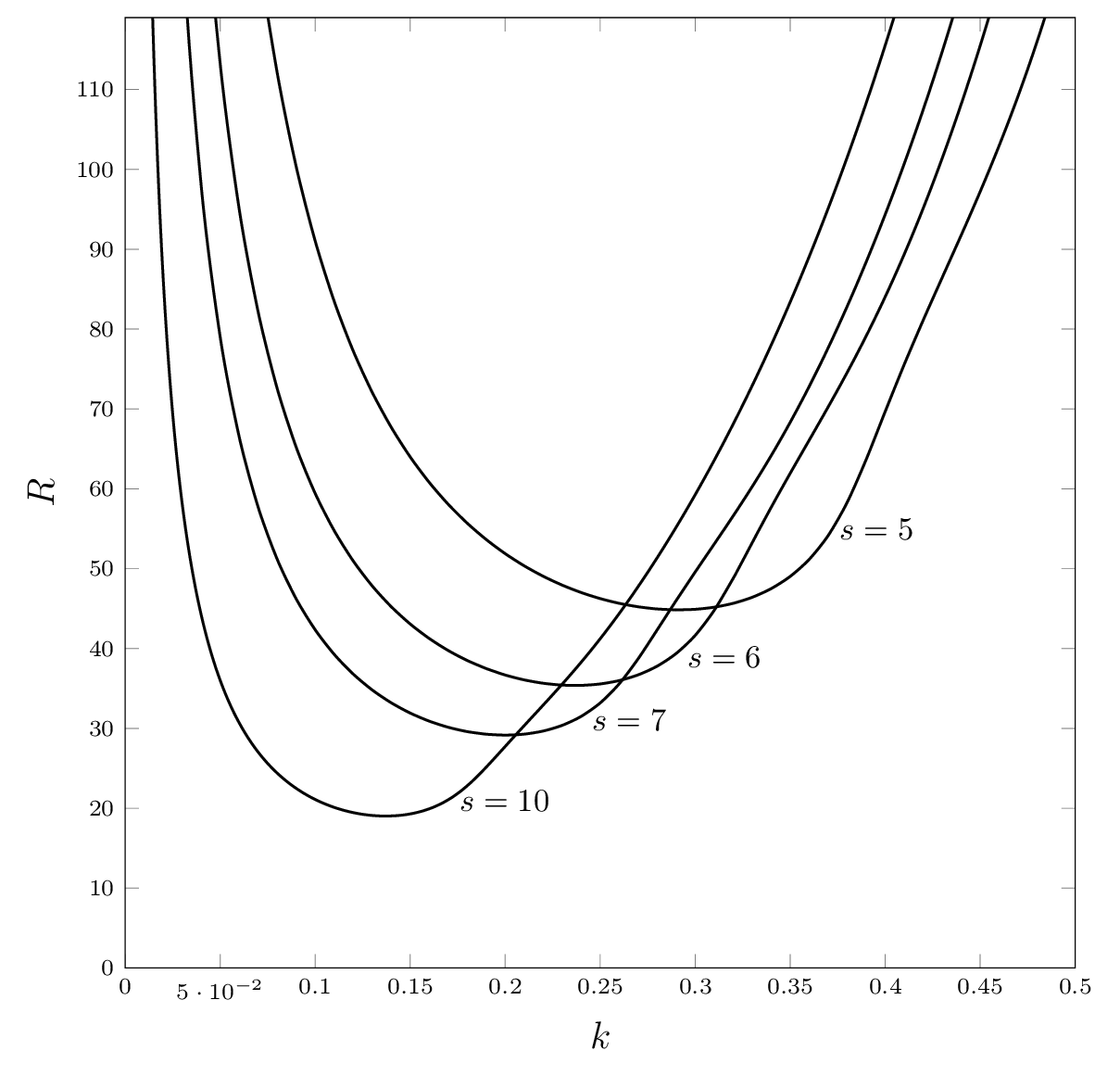}
\caption{Neutral stability curves in the $(k, R)$ parametric plane for axisymmetric modes $(m=0)$, with $s=5,\ 6,\ 7$ and $10$}
\label{fig6}
\end{figure}

\subsection{Neutral stability}

The study of the neutral stability at larger aspect ratios is carried out in Figs.~\ref{fig5} and \ref{fig6} for values of $s$ greater than $2$. There is an evident general trend showing that the onset of linear instability happens with smaller and smaller values of $R$ as $s$ increases, and that the normal modes triggering the instability have gradually smaller values of $k$ for increasing $s$. In particular, Fig.~\ref{fig5} illustrates cases where $2 \leqslant s \leqslant 4$ and compares the behaviour of axisymmetric modes $(m=0)$ with that of the lowest non--axisymmetric modes $(m=1)$. In all cases, the axisymmetric modes yield the least threshold to the instability. With this understanding, the curves reported in Fig.~\ref{fig6} for larger aspect ratios, {\em i.e.} $5 \leqslant s \leqslant 10$ are relative to axisymmetric modes. 
We see that the normal modes activating the instability {correspond to decreasing wavenumbers} as $s$ increases, which physically means very large wavelengths. 
{More precisely, the wavelength of the cells is roughly the same as the radial width of the annular slab.}

\begin{figure}[t!!]
\centering
\includegraphics[width=\textwidth]{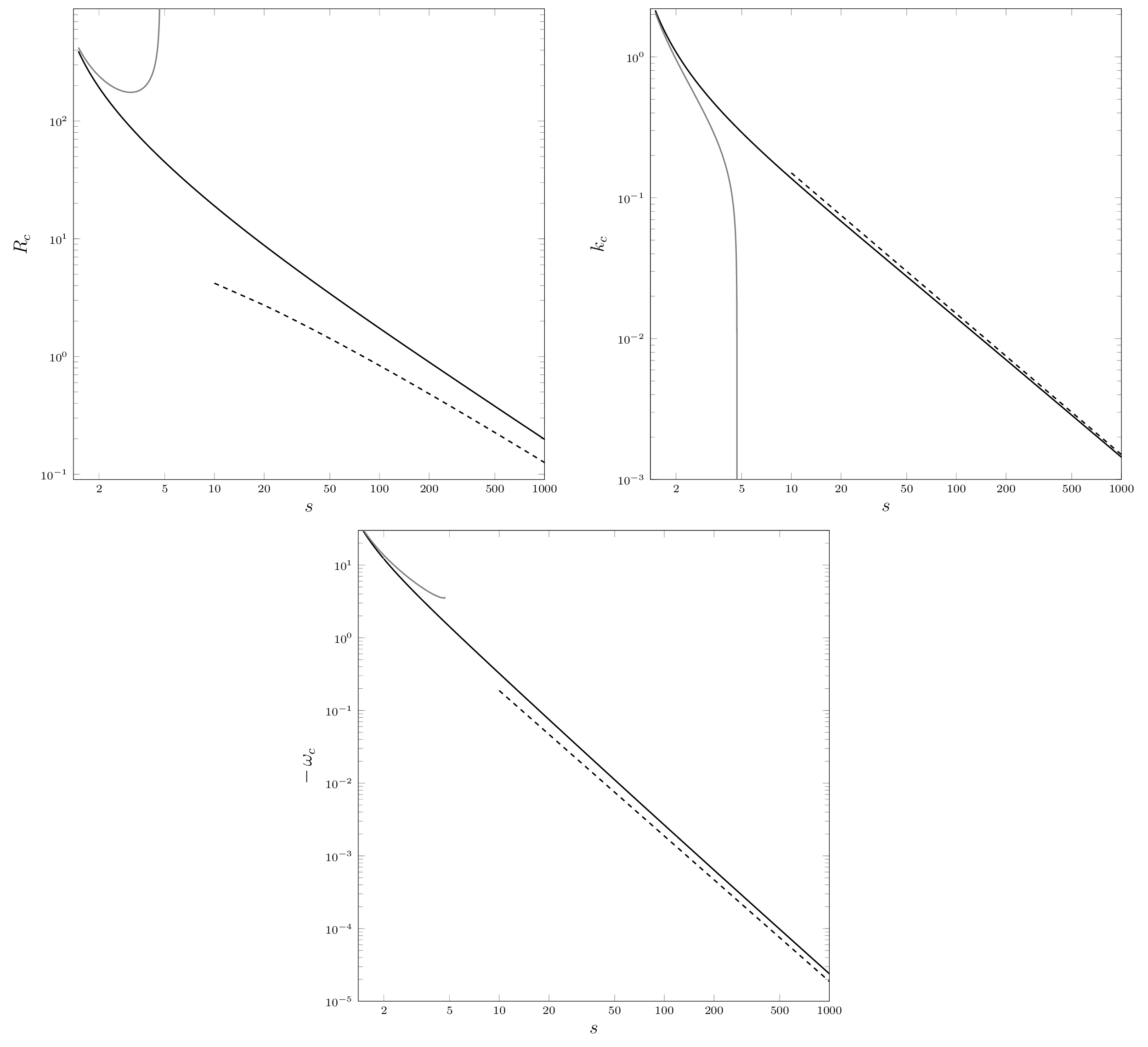}
\caption{Critical values of $R$, $k$ and $\omega$. Solid black lines are for axisymmetric modes $(m=0)$. Solid gray lines are for $m=1$ modes. Dashed black lines denote the asymptotic solution for $s \gg 1$ and $m=0$}
\label{fig7}
\end{figure}

\begin{table}[t]
\begin{center}
\begin{tabular}{l|l|l|l}
$\ s$ & $\quad R_c$ & $\quad\ \ k_c$ & $\qquad \omega_c$\\
\hline\hline
 & & \\[-3mm]
1.5 & 390.893 & 2.13600 & $-29.5279$ \\
1.7 & 277.626 & 1.53411 & $-19.4465$ \\
2   & 192.558 & 1.08351 & $-12.1651$ \\
3   & 93.4923 & 0.557727 & $-4.45151$ \\
4   & 60.8820 & 0.380902 & $-2.32196$ \\
5   & 44.8444 & 0.291283 & $-1.42359$ \\
6   & 35.3772 & 0.236747 & $-0.960085$ \\
7   & 29.1588 & 0.199898 & $-0.690001$ \\
10  & 19.0241 & 0.137167 & $-0.323306$ \\
100 & 1.73774 & 0.0139895 & $-0.00265745$ \\
\end{tabular}
\end{center}
\caption{\label{tab2}Critical values of $R$, $k$ and $\omega$ for $m=0$ and different aspect ratios $s$}
\end{table}%

Figure~\ref{fig7} displays plots of $R_c$, $k_c$ and $\omega_c$ versus $s$ in the range $1.5 \leqslant s \leqslant 10^3$. Both the branch of axisymmetric modes, $m=0$, and the lowest branch of non--axisymmetric modes, $m=1$, are reported. One may see that axisymmetric modes {have the lowest} onset conditions for instability in the whole range considered in Fig.~\ref{fig7}. In fact, the value of $R_c$ for a given aspect ratio is smaller for $m=0$ than for $m=1$. We also note that the modes $m=1$ show a vertical asymptote for both $R_c$ and $k_c$ when $s$ attains approximately the value $4.69$. Close to this asymptote, $R_c$ tends to infinity, while $k_c$ tends to zero. On the other hand, $\omega_c$ approaches a finite value, $-3.54$. For $s > 4.69$, no unstable $m=1$ modes are detected.

The general trend for axisymmetric modes {is} that an increasing aspect ratio $s$ leads to {an increasingly} unstable situation as $R_c$ is a monotonic decreasing function of $s$. When $s \to \infty$, we have $R_c \to 0$. This feature is evident from the asymptotic analysis presented in the forthcoming Section~\ref{asy}. The results of this asymptotic solution are employed to try to capture the trend for $R_c$, $k_c$ and $\omega_c$ when $s \gg 1$, with the asymptotic data represented by the dashed lines in Fig.~\ref{fig7}. In the range reported in Fig.~\ref{fig7}, we reckon that the asymptotic solution does not match accurately the trend of $R_c$ versus $s$, while a much better approximation is found for $k_c$ and for $\omega_c$ versus $s$.

Table~\ref{tab2} reports the critical values $R_c$, $k_c$ and $\omega_c$ corresponding to some sample aspect ratios $s$ ranging from $1.5$ to $100$. These data are excerpted from those employed for the plots in Fig.~\ref{fig7}.

\subsection{Asymptotic solution for $s \to \infty$}\label{asy}
In order to capture the trend {for} the neutral stability data when $s$ is extremely large, we look for an asymptotic solution. Our focus will be on the axisymmetric modes which turned out to be {the most dangerous modes}
so that we will fix $m=0$. We mention that the limit $s \to \infty$ means a situation where the annular cross--section ($z=constant$ domain) tends to the semi--infinite region $r \geqslant 1$. We thus define the rescaled quantities
\eqn{
y = - \ln\!\left( \frac{r}{s} \right), \quad k_0 = k\,s, \quad \lambda_0 = \lambda\, s^2, \quad R_0 = R\; \frac{s}{\ln(s)}, \quad h_0 = h\, \ln(s)  .
}{25}
One can easily recognise that $y=0$ identifies the external boundary $(r=s)$, while the internal boundary $(r=1)$ {corresponds to}
$y=\ln(s)$.

\begin{figure}[t!!]
\centering
\includegraphics[width=0.7\textwidth]{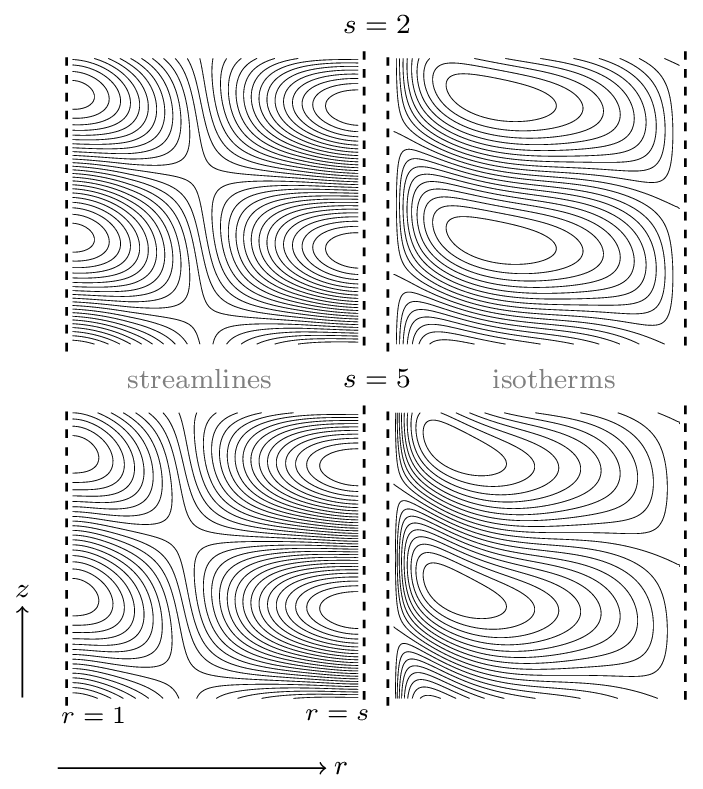}
\caption{Streamlines $(\hat\psi = constant)$ and isotherms $(\hat T=constant)$ for axisymmetric modes $(m=0)$ at the onset of the instability $(R=R_c,\ k=k_c,\ \omega=\omega_c)$, with $s=2$ and $s=5$}
\label{fig8}
\end{figure}

Then, by taking the limit $s \to \infty$, we can rewrite \equas{15} as
\gat{
\adervso{f}{y} - e^{-2 y}\left( k_0^2\, f + i\,k_0\, R_0\, h_0  \right) = 0 , \label{26a}\\
\adervso{h_0}{y} - e^{-2 y}\left[ k_0^2 + i\, k_0\, R_0 \left( \frac{1}{2} - y \right) + \lambda_0 \right] h_0 - \adervo{f}{y} = 0, \label{26b}\\
y=0,\ y\to +\infty : \qquad f = 0, \quad h_0 = 0. \label{26c}
}{26}
Equations~(\ref{26}) define an eigenvalue problem over the semi--infinite range $0 \leqslant y < +\infty$. Its solution can be found by adopting the same technique described in Section~\ref{solmed}. The initial point for the numerical solver is $y=0$, the end point conditions are now formulated as limiting conditions attained when $y \to +\infty$. This circumstance complicates the application of the shooting method. In fact, one has to track the change of the computed eigenvalue as the end point migrates to gradually larger values of $y$. This tracking process is interrupted when a prescribed number of significant figures of the eigenvalue remains unaltered by a further increment of the end point position, $y$. 
By this solution strategy, we are able to compute the critical values of $R_0$, $k_0$ and $\omega_0$ with three digits accuracy. They are given by
\eqn{
R_{0c} = 18.2, \quad k_{0c} = 1.50, \quad \omega_{0c} = - 18.8 .
}{27}
On account of \equasa{25}{27}, asymptotic expressions for $R_c$, $k_c$ and $\omega_c$ when $s \gg 1$ can be written as
\eqn{
R_{c} = 18.2\; \frac{\ln(s)}{s}, \quad k_{c} = \frac{1.50}{s}, \quad \omega_{c} = - \frac{18.8}{s^2} .
}{28}
An evident feature emerging from \equa{28} is that $R_c$, $k_c$ and $\omega_c$ tend to zero when $s \to \infty$, which is expected from the graphical trends exploited in Fig.~\ref{fig7}.

\begin{figure}[t!!]
\centering
\includegraphics[width=0.7\textwidth]{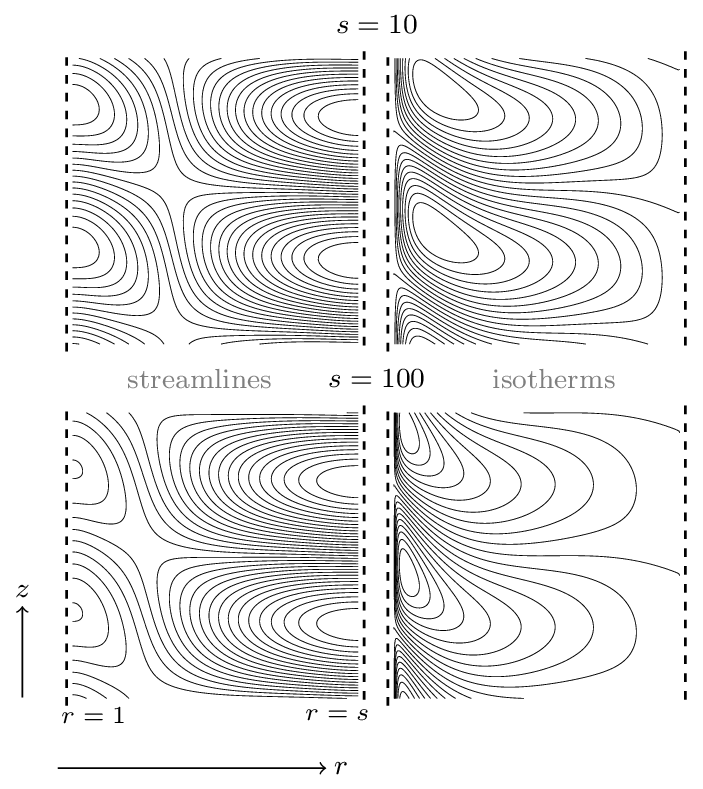}
\caption{Streamlines $(\hat\psi = constant)$ and isotherms $(\hat T=constant)$ for axisymmetric modes $(m=0)$ at the onset of the instability $(R=R_c,\ k=k_c,\ \omega=\omega_c)$, with $s=10$ and $s=100$}
\label{fig9}
\end{figure}

\subsection{Streamlines and isotherms}
Displaying the streamlines and isotherms of the perturbation modes triggering the instability {always provides} important information {such as mode shapes, the thickness of the disturbance field relative to that of the basic state and it often} guides further investigations on the topic. For {example,} the shape of the convection cells {obtained from linearised stability theory}
may be the starting point for analysing the phenomenon of nonlinear saturation that is expected to happen at moderately supercritical Rayleigh numbers \citep{barletta2015proof}.

In order to define the streamfunction we first recall that the most unstable modes are those with $m=0$, {\em i.e.} the axisymmetric modes. For such modes, \equa{11c} yields $\hat v=0$. Then, \equa{11a} can be rewritten as
\eqn{
\aderv{\left(r\, \hat{u}\right)}{r} + \aderv{\left(r\, \hat{w}\right)}{z} = 0.
}{29}
Equation~(\ref{29}) is identically satisfied by defining the streamfunction, $\hat{\psi}$, as
\eqn{
\hat{u} = \frac{1}{r}\; \aderv{\hat \psi}{z}, \quad \hat{w} = -\frac{1}{r}\; \aderv{\hat \psi}{r}.
}{30}
We now utilise \equasa{11b}{30} to write
\eqn{
\aderv{\hat p}{r} = - \frac{1}{r}\; \aderv{\hat \psi}{z}.
}{31}
This equation is satisfied, on account of \equa{14}, by expressing the streamfunction as
\eqn{
\hat{\psi}(r,z,t) = i\; \frac{r}{k}\; f'(r)\, e^{i(k z - \omega t)}\, e^{\eta t}.
}{32}
By evaluating the real part of the right hand side of \equa{32}, one can obtain the $\hat \psi = constant$ lines, namely the streamlines. Just the same can be done by computing the real part of $\hat T$, through \equa{14}, in order to obtain the isotherms of the normal mode perturbations. Drawing the streamlines and the isotherms is specially significant at the onset of the instability, {\em i.e.} at the critical conditions $R=R_c$, $k=k_c$ and $\omega = \omega_c$. Plots of the streamlines and isotherms at critical conditions are provided in Figs.~\ref{fig8} and \ref{fig9} for the aspect ratios $s=2,\ 5$ and $s=10,\ 100$, respectively. Figures~\ref{fig8} and \ref{fig9} highlight the increasing asymmetry of the convection cells with respect to the radial midpoint, $r = (1+s)/2$, as the aspect ratio $s$ increases. In fact, the midpoint symmetry of the cellular patterns is a feature of the plane slab case, pointed out by \citet{barletta2015proof}. Figures~\ref{fig8} and \ref{fig9} show that, with larger values of $s$, the streamlines and isotherms tend to be compressed close to the inner radial boundary, $r=1$, and stretched in the rest of the domain. We mention that, in the captions of these figures, time is unspecified (or, equivalently, it is set to $t=0$) as the value of $t$ just yields a phase shift of the patterns along the $z$ direction, and an overall scale of the real parts of $\hat \psi$ and $\hat T$. Both these features are barely important for such plots. Finally, we point out that the scale of the $z$ coordinate is adjusted in order to adapt the vertical domain to a full wavelength $(2\pi/k)$ whatever is the aspect ratio $s$.

The dramatic increase of the temperature gradient close to $r=1$ when the aspect ratio $s$ is very high, as documented in Fig.~\ref{fig7}, sheds light upon the extremely slow convergence experienced when solving numerically the eigenvalue problem (\ref{26}). We recall that, as it is shown in \equa{27}, we were able to compute $R_{0c}$, $k_{0c}$ and $\omega_{0c}$ with three significant figures.

\section{Conclusions}
The stability of the stationary parallel flow in a vertical cylindrical porous slab has been analysed. We investigated a basic flow induced by the thermal buoyancy due to the temperature difference between the bounding cylindrical surfaces. Indeed, the basic state features a purely horizontal temperature gradient oriented radially. The bounding radial surfaces are assumed to be permeable and open to external fluid environments maintained in a rest state. A linear stability analysis of the basic state has been performed by employing a normal mode formulation of the eigenvalue problem which involves the Rayleigh number, $R$, and the aspect ratio, $s$, between the external and internal radii of the bounding surfaces. The numerical solution of the resulting eigenvalue problem served to obtain the neutral stability condition and the critical value of $R$, for different aspect ratios $s$. The main findings for our analysis are summarised as follows:

\begin{itemize}
\item The most unstable normal modes at the onset of instability are axisymmetric. Normal modes invariant along the axial direction turned out to be always stable. Thus, the evaluation of the neutral stability condition could be carried out through a two--dimensional solution of the governing equations.
\item In the asymptotic case $s\to 1$, the neutral stability curve and, hence, the critical values of $R$, of the wavenumber $k$ and of  the angular frequency $\omega$ match exactly those obtained for a plane vertical slab with the same boundary conditions. This case was extensively investigated in a previous study \citep{barletta2015proof}. In order to address properly this limit, an alternative definition of the Rayleigh number has been employed, based on the slab thinckness as the reference length, instead of the internal radius of the annulus. 
\item The curvature of the slab entails a destabilisation of the basic flow. The critical Rayleigh number, $R_c$, decreases monotonically with an increasing aspect ratio $s$. In the limit $s \to \infty$, $R_c$ tends to zero or, equivalently, the basic flow is always unstable. This limit addresses a situation where the porous medium is semi--infinite and internally bounded by a cylindrical surface.
\item An increasing aspect ratio $s$ causes an unbounded increase of the temperature gradient close to the internal cylindrical boundary. This phenomenon has been displayed for the perturbing normal modes associated with the critical values of $R$, $k$ and $\omega$. An asymptotic solution of the eigenvalue problem valid for $s \gg 1$ has been obtained.
\end{itemize}

There are several opportunities for future developments of this research. The supercritical regime and the emergence of a mechanism of nonlinear saturation is a definitely interesting task which can be approached either with a weakly nonlinear analysis of instability or with a fully nonlinear numerical computation. Such approaches can also allow an inspection of the possible existence of a subcritical bifurcation. To achieve this goal, the energy method can be an alternative invaluable tool for exploring the dynamics of finite amplitude disturbances acting upon the basic flow state. Another aspect that will be explored in a separate paper is the extension of Gill's no--go theorem for the instability \citep{gill1969proof} when the bounding cylindrical surfaces are modelled as impermeable.

\section*{Acknowledgements}
The authors A. Barletta and M. Celli acknowledge the financial support from the grant PRIN 2017F7KZWS provided by the Italian Ministry of Education and Scientific Research.

\bibliography{biblio}
\bibliographystyle{model1-num-names}

\end{document}